\documentclass[sigconf]{acmart}

\usepackage{booktabs}
\usepackage{listings}
\usepackage{tabularx}
\usepackage{array}

\setcopyright{rightsretained}
\usepackage{xcolor}

\acmDOI{}

\acmConference[KDD 2017]{23rd ACM SIGKDD International Conference on Knowledge Discovery and Data Mining, Data-Driven Discovery Workshop}{August 2017}{Halifax, Nova Scotia, Canada} 
\acmYear{2017}
\copyrightyear{2017}

\acmPrice{}

\begin{document}
\title[Marve: Measurement Context Extraction from Text]{Measurement Context Extraction from Text: Discovering Opportunities and Gaps in Earth Science}


\author{Kyle Hundman\textsuperscript{1}, Chris A. Mattmann\textsuperscript{1,2}}
\email{{kyle.a.hundman,chris.a.mattmann}@jpl.nasa.gov}
  \affiliation{\begin{tabular}{*{2}{>{\centering}p{.45\textwidth}}}\textsuperscript{1}Jet Propulsion Laboratory & \textsuperscript{2}Computer Science Department \tabularnewline California Institute of Technology & University of Southern California \tabularnewline Pasadena, CA 91109 USA & Los Angeles, CA 90089 USA \end{tabular}}

\renewcommand{\shortauthors}{K. Hundman, C. Mattmann}

\begin{abstract}
We propose Marve, a system for extracting measurement values, units, and related words from natural language text. Marve uses conditional random fields (CRF) to identify measurement values and units, followed by a rule-based system to find related entities, descriptors and modifiers within a sentence. Sentence tokens are represented by an undirected graphical model, and rules are based on part-of-speech and word dependency patterns connecting values and units to contextual words. Marve is unique in its focus on measurement context and early experimentation demonstrates Marve's ability to generate high-precision extractions with strong recall. We also discuss Marve's role in refining measurement requirements for NASA's proposed HyspIRI mission, a hyperspectral infrared imaging satellite that will study the world's ecosystems. In general, our work with HyspIRI demonstrates the value of semantic measurement extractions in characterizing quantitative discussion contained in large corpuses of natural language text. These extractions accelerate broad, cross-cutting research and expose scientists new algorithmic approaches and experimental nuances. They also facilitate identification of scientific opportunities enabled by HyspIRI leading to more efficient scientific investment and research. 
\end{abstract}

%
%




\maketitle

\section{Introduction}

Much of the world's scientific information is easily accessible from our fingertips. For example, a search for ``remote sensing'' on Thomson Reuter's Web of Science yields nearly 14,000 journal articles from 2014--2016\footnote{\url{https://apps.webofknowledge.com}}. However, the careful analysis and understanding of that scientific information is not easy. An individual scientist may spend many hours (re-)reading a single article to comprehend its message and significance. The aforementioned corpus of remote sensing articles is simply too much information for a human, or even a small set of them, to read and synthesize into knowledge. Fortunately, continual advances in search and natural language processing (NLP) have greatly enhanced our ability to automatically characterize and sift through large-scale unstructured data. Neural network approaches have vastly improved essential NLP tasks such as part-of-speech (POS) tagging and dependency parsing. For example, Stanford CoreNLP's dependency parser achieved 91.7\% \cite{chen2014fast} accuracy on the Penn Treebank dataset and Google's Parsey McParseface attained an impressive 94.4\% on sentences from various news sources \cite{andor2016globally}. 

While these foundational NLP tasks provide a framework for processing core textual components, deriving scientific meaning and nuance from those components is more complex. One of the core elements of science is {\em measurement}, which involves quantification (in units such as nanometers or kilograms), situational context (e.g. location, timeframe, sentiment) and often statistical modifiers (e.g. average). Consider the sentence, ``The unexpected drop in stratospheric water vapor slowed the rate of increase in surface temperature in the subsequent decade by 25\%.'' Identifying ``25\%'' as modifying the ``rate of increase in surface temperature'' is difficult without a system that considers the underlying structure of the sentence. Proper measurement extraction and labeling enables the creation of unique knowledge bases and opens exciting possibilities for modeling and visualization techniques that rely on organized and uniform numerical data. Automatically discerning scientific measurements and specific contextual aspects can allow for quick summarization and scientific understanding of a large corpus of literature and/or news articles. In addition, it can allow for validation and comparison of automatically extracted and categorized measurements with raw measurement data; this can provide additional context and even scientific corroboration of phenomena. 

We describe a framework, Marve, that fuses and extends existing techniques in NLP and text processing to extract context around measurements in natural language. Using rules primarily based on word dependencies and POS tags, Marve exploits a limited set of approximately 10 parts of speech and 15-20 word dependency types and English language patterns used to describe measurements and the objects or concepts they quantify. Traditionally, the cost of manual curation and the ambiguity of unaccompanied measurements have limited the collection and application of semantic measurement data. Marve circumvents these problems by understanding contexts and consequently improving identification of measurement types. From a scientific perspective, Marve accelerates exploration of literature and promotes cross-pollination of ideas and approaches across domains. 

\section{Motivation}\label{sec:motivation}
Marve originated from a NASA Advanced Concepts project that has provided data-driven support for NASA's proposed Hyperspectral Infrared Imager (HyspIRI) mission, which will monitor a variety of ecological and geological features at a wide range of wavelengths. The planned HyspIRI instrumentation has unique technical capabilities such as high spatial resolution and hyperspectral coverage that will benefit several scientific areas \cite{lee2015introduction}. However, the extent and nature of these benefits are not easily understood because much of this information is embedded within scientific publications spread across numerous journals. We set out to automatically identify and profile these new scientific opportunities using a corpus of approximately 2,500 recent publications and abstracts from various journals in the remote sensing domain. 

Our first approach involved the use of regular expressions to extract common measurement types in the remote sensing domain. Extracted measurements like {\em spatial resolution}, {\em spectral coverage} and {\em revisit rate} provided a useful bookmarking of our corpus -- discussion around hyperspectral wavelengths (\textgreater2.4 $\mu m$) and high spatial resolutions were pointers toward potential science enabled by HyspIRI. Visualizing these extractions also revealed the scale of the discussion around various wavelengths (as shown in Figure 1). 

\begin{figure}
\includegraphics[height=2.4in]{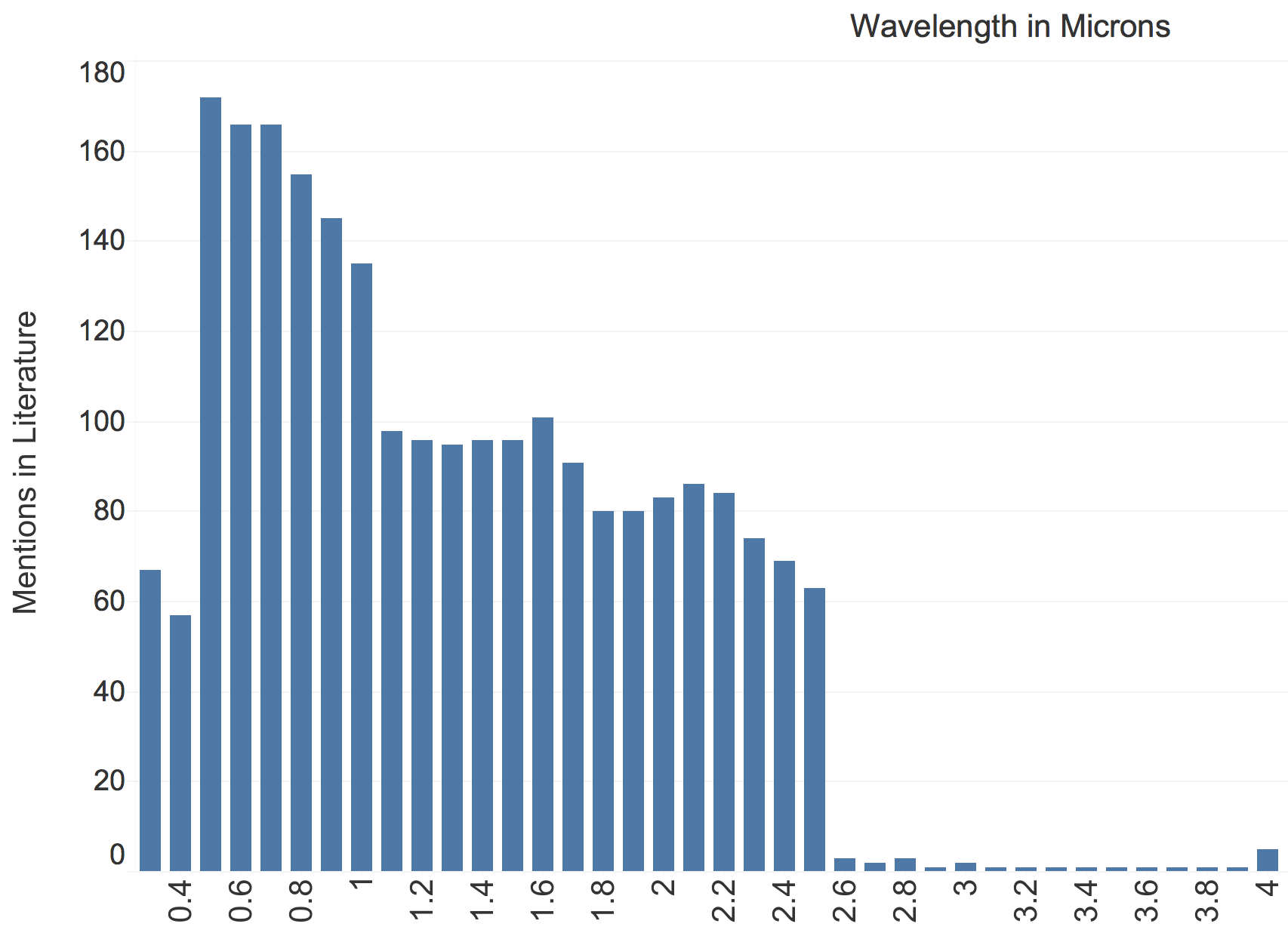}
\caption{A histogram of extracted spectral wavelengths from the corpus of remote sensing publications and abstracts used for HyspIRI. Most of the discussion takes place around visible and near infrared wavelengths ($0.4\ to \ 2.5\mu m$), but opportunities for new or improved science enabled by HyspIRI hysperspectral coverage may most likely be found in papers discussing longer-wave infrared wavelengths (e.g. $4\mu m$ on the right side of the chart. (credit: Jason Hyon)}
\end{figure}

Unfortunately, regular expressions didn't generalize across measurement types, precision and recall of extractions was unknown, and regular expressions are complex. Most importantly, measurements were extracted in isolation. An extraction of ``50 m'' could be a measurement of height, swath, length, or resolution in the context of remote sensing. Ultimately, Marve resulted from our pursuit of a general, accurate, and precise tool that provided semantically-rich extractions.

\section{Related Work}\label{sec:rel}
\subsection{Measurement Extraction}
\subsubsection{Grobid Quantities}
The Marve stack includes Grobid Quantities \cite{lopez2010automatic}, a library that utilizes linear conditional random fields (CRF) to identify measurement units and values. Training the Grobid model requires labeled training data, ideally from the target domain. And although labeled data is costly, supervised machine learning is appropriate for measurement extraction; measurement conventions and unit formats vary widely across scientific domains and the resulting proliferation of patterns is too large and varied for unsupervised models or rule-based systems to be effective. We initially tried extracting measurements using a rule-based approach built around part of speech (PoS) tags, named-entity recognition (NER), word dependencies, and regular expressions. While this method was fast and training-free, attempts at generalizing the system for different domains led to numerous false positive extractions.

Through this experimentation we determined that quantified substances and related entities (i.e. context) don't present the same challenges as measurement values and units. Instead, they follow common language patterns that generalize well. This allows Marve to identify words and entities related to a measurement without labeled examples and model training. Marve can also capture related entities from a broad assortment of language patterns. Consider the following sentence: ``Landsat-8 achieved 82\% classification accuracy for cutleaf teasal.'' Grobid Quantities isn't designed to identify ``Landsat-8'' or ``cutleaf teasal'' as related entities. Marve is able to capture this additional information without domain-specific labels or training -- these types of phrasal patterns and clauses are common across the English language. Grobid Quantities could be extended to capture more context, but tuning its extraction process requires adding or adjusting labels and re-training the full model for a specific domain. Marve mitigates additional overhead required for context extraction and uses Grobid Quantities as an off-the-shelf dependency.

\subsubsection{Quantalyze and GATE}
Quantalyze\footnote{https://www.quantalyze.com/} is a commercial product that also performs measurement, unit, and context extraction. Evaluation of its performance can only be achieved through their online demo, but after comparing extractions from several paragraphs of text, their tool appears to achieve poor recall in both measurements captured and quantified substances captured. 

Additionally, Agatonovic et. al \cite{agatonovic2008large} employ the General Architecture for Text Engineering (GATE) \cite{cunningham2002framework} to extract measurement values and units from patent documents. Their approach involves building patent-specific gazateers and hand-written rules to generate measurement annotations using the GATE framework. While Grobid Quantities requires labeled data, its embedded CRF  model obviates rule-writing and Agatonovic et. al's approach only captures accompanying words such as ``less than'' or ``between'' while ignoring related entities and context.

\subsection{Open Information Extraction (OIE)}
Various OIE systems approach relation extraction in similar ways to Marve. KrakeN \cite{akbik2012kraken}, CSD-IE \cite{bast2013open} and ClausIE \cite{del2013clausie} utilize dependency patterns and POS tags to detect clauses and find their propositions. This information is then used to construct triples representing facts in a corpus, such as: (``Kelly'', ``finished'', ``nursing school''). Similar to the Agatonovic et. al's GATE-based approach, certain measurements could be extracted via OIE approaches. But OIE is centered on verb-mediated propositions and measurement context occurs in a variety of other forms such as adverbials: ``The satellite captured imagery with 50 m spatial resolution.'' This leads to poor OIE recall for measurements. Also, when measurements are extracted by these systems the output requires significant post-processing to filter extraneous OIE extractions and properly separate measurements, units, and related entities. Marve is a more directed form of these approaches, and it is most similar to those that sacrifice efficiency for improved precision and recall (e.g. OLLIE \cite{schmitz2012open}, Kraken, and ClausIE). 

\subsection{Relationship Extraction and Knowledge Base Construction}
Knowledge base construction (KBC) and relationship extraction have migrated from pattern matching and rule-based systems to machine learning based systems over the last decade. One of the driving factors behind this trend is that KBC systems that rely on a multitude of rules require some assurance of the precision and recall of such rules \cite{niu2012deepdive}. In practice this is difficult and tedious.  As discussed in Section 3.1, this is also the reason for the use of machine learning approaches to measurement value and unit extraction - too many patterns and rules result in uncertainties about precision and recall. However, Marve differs from traditional KBC systems in two primary ways. First, Marve does not explicitly classify types of relationships between extracted measurements and their related words and entities. Second, Marve is directed at a very specific type of extraction (measurements) that benefits many scientific information extraction scenarios. In this sense, it is complementary to broader KBC systems.  One of the most prominent KBC systems of late is DeepDive, which is designed to identify relationships between extraction types using labeling functions written by domain experts. However, generating the extractions of interest is left up to the user and writing discerning labeling functions is a gradual, iterative process. Marve automates a large part of the development of a DeepDive system by automatically extracting measurements and providing precise measurement context to labeling functions. Section 6.2 includes further discussion around Marve and DeepDive integrations.

\section{Methodology}
\subsection{Overview}
As discussed in Section ~\ref{sec:rel}, we decided against a custom measurement extractor and instead used Grobid Quantities to extract measurement values and units. Like Marve, Grobid Quantities represents sentences with undirected graphs. Though instead of parsing language patterns, Grobid uses a probabilistic graphical CRF method that learns parameter values through maximum likelihood estimation. This approach to extracting numerical values and units was more consistent in our experimentation although it adds processing overhead and requires labeled training data.

Once measurements values and units are identified using Grobid, Stanford's CoreNLP library \cite{manning-EtAl:2014:P14-5} is used to perform more traditional NLP tasks such as tokenization, PoS tagging, and word dependency parsing. Marve uses combinations of the output from these tasks to identify measurement types (e.g. 10 m \textit{spatial resolution}) and related entities and descriptors (e.g. \textit{Hannibal} had \textit{around} 40 elephants). When represented in a graph (see Figure 2), these patterns originate at the measurement unit token(s) and expand outward to connected nodes (words). If Marve finds a pattern defined as valid its pre-defined rules (represented using JSON\footnote{\url{https://github.com/khundman/marve/blob/master/marve/dependency_patterns.json}}), the resulting word(s) will be returned as  related to the measurement. 

\begin{figure*}
\includegraphics[height=3.0in]{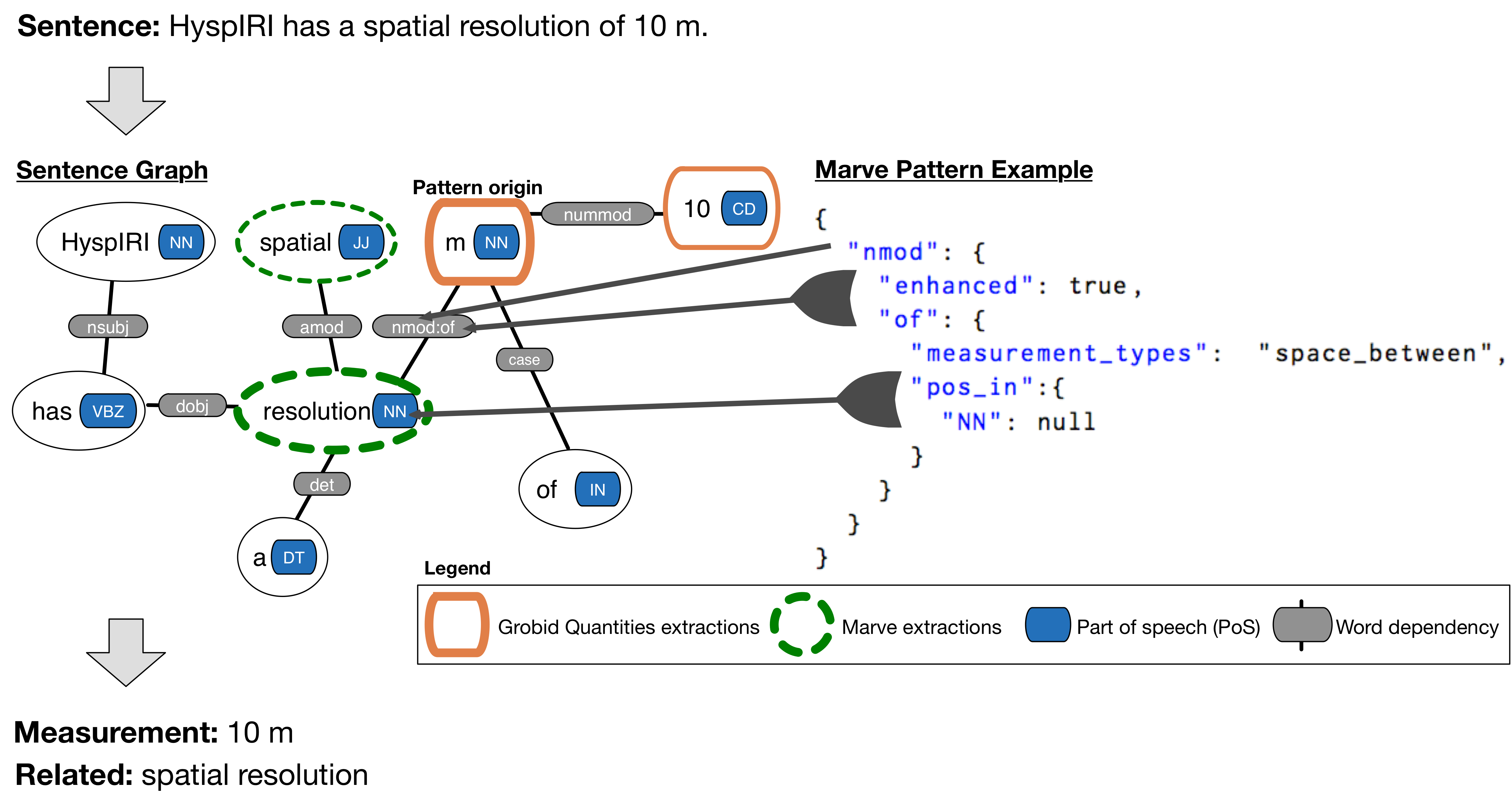}
\caption{An example detailing the measurement and context extraction process. Word dependencies and PoS tags are loaded into a graph, where patterns defined in Marve are evaluated to identify additional context (e.g. measurement types).}
\end{figure*}

\subsection{Model Structure}
Consider a connected, undirected graph $G = (V, E)$ where $V$ and $E$ denote the sets of nodes and edges respectively, such that: \begin{itemize}
  \item $S = \{s_{1}, s_{2},..., s_{n}\}$ is a set of sentences that comprise a corpus of text from which  measurements are extracted.
    \item $T = \{t_{1}, t_{2},...,t_{n}\}$ is a set of all tokens in $S$.
      \item $s_{i} = \{t\mid t \in T\}$ where each sentence $s \in S$ is a set of t tokens.
        \item One graph $G$ is constructed for each sentence $s_{i}$.
          \item $L = \{l_{1}, l_{2},...,l_{n}\}$  where $l_{i}$  is a label which identifies the part of speech for each $t_{i}$ token in a sentence $s_{i}$.
\end{itemize}
Given these notations, the set of nodes $V$ in each graph can be defined as $V = \{v_{1}, v_{2},...,v_{n}\}$ where $v_{i}$  stores a token $t_{i} \in t$ for a set of $t$ tokens in a sentence $s$, and each token $t_{i}$ is labeled with label $l_{j} \in L$. Then we define $e_{ij}$ as an edge connecting $(v_{i}, v_{j})$ with a label $d_{ij}$ representing the dependency between tokens $(t_{i}, t_{j})$, where $D$ is the set of all dependencies equal to the length of $E$.

\subsection{Pattern Matching}

Once a measurement is identified by Grobid Quantities, the token $t_{i}$ that corresponds to the measurement unit becomes the origin for subsequent pattern evaluation (as shown by the thicker circle around ``m'' in Figure 2), and a graph $g_{i}$ is constructed for sentence $s_{i}$ containing the measurement. Evaluation continues if the dependency label $d_{ij}$ for any edge $e_{ij}$ originating at $t_{i}$ matches the word dependency types defined in the valid dependency patterns. One subtlety stems from CoreNLP's enhanced dependencies, which provide the connecting word for certain dependency types. For example, if the conjunction ``and'' connects two words, the enhanced dependency type returned by CoreNLP is ``conj:and'' rather than ``conj.'' When Marve encounters a dependency type that has been enhanced, it evaluates the enhanced portion separately allowing for more nuanced pattern definitions. This is represented in the JSON structure with a boolean value for the ``enhanced'' key (shown in Figure 2). If ``enhanced'' is true, the dependency label $d_{ij}$ is split into two parts: the connecting word and the dependency type. If both parts match the JSON structure, evaluation continues along that nested path.

\lstset{
    basicstyle=\footnotesize\ttfamily,
    string=[s]{"}{"},
    stringstyle=\color{blue},
    comment=[l]{:},
    commentstyle=\color{black},
    showstringspaces=false
}

\begin{figure}
  \caption{Sample Marve output for the example sentence in Figure 2. The ``quantity'' field is populated by Grobid Quantities and the ``related'' field is added by Marve.}
  \label{tab:out}
\begin{lstlisting}
{
  "type": "value",
  "quantity": {
    "parsedValue": 10,
    "normalizedQuantity": 10,
    "rawValue": "10",
    "rawUnit": {
      "offsetStart": 39,
      "offsetEnd": 40,
      "tokenIndices": ["8"],
      "name": "m"
    },
    "offsetEnd": 38,
    "offsetStart": 36,
    "tokenIndex": 7,
    "normalizedUnit": {
      "type": "length",
      "name": "m",
      "system": "SI base"
    },
    "type": "length"
  }
  "related": [
    {
      "rawName": "resolution",
      "connector": "",
      "offsetEnd": 32,
      "relationForm": "nmod:of",
      "offsetStart": 22,
      "tokenIndex": 5,
      "descriptors": [
        {
          "rawName": "spatial",
          "tokenIndex": "4"
        }
      ]
    }
  ]
}
\end{lstlisting}
\end{figure}

Marve first considers the format of the measurement after word dependencies are evaluated. Three primary formats are defined:
\begin{itemize}
\item \texttt{space\_between} (e.g. ``10 m'')
\item \texttt{attached} (e.g. ``10m'')
\item \texttt{hyphenated} (e.g. ``10-m'')
\end{itemize}
Measurement formats are identified using the character indices of measurement value token $t_{k}$ and measurement unit token $t_{i}$. If $t_{i}$ and $t_{k}$ are adjacent (without a space), they are ``attached.'' If not, a simple check for a space or hyphen is performed. Word dependency and PoS patterns vary based on these formats and explicitly defining rules for them improves Marve's precision. 

PoS tags are the next evaluation step in the Marve system. If the measurement format is valid according to the dependency pattern and token $t_{j}$ is also connected to the unit token $t_{i}$ via a valid dependency pattern, label $l_{j}$ is evaluated in one of two ways:
\begin{itemize}
\item \texttt{pos\_in}: As long as one of the keys in the \texttt{pos\_in} value matches part of label $l_{j}$ they are valid (e.g. if one of the keys for the \texttt{pos\_in} nested object is ``NN'' and label $l_{j}$  is ``NNS'')
\item \texttt{pos\_equals}: The specified PoS labels must match label $l_{j}$ exactly
\end{itemize}
If a matching PoS key has its own keys and values in the JSON, it is considered a special case. Most of these cases involve verbs, which are often part of a clause containing a subject related to the measurement. If so, all nodes connected to token $t_{j}$ are evaluated by a separate function. Valid word dependencies are passed as parameters to this function, and it executes recursively if it encounters additional connected verb tokens. If the value of a matching PoS key in the JSON is null, no more evaluation is needed and the token $t_{j}$ is returned as a related word. 

The last step in constructing the output is finding adjectives, modifiers, or compounds connected to related nouns. This includes words like ``spatial'' in the example in Figure 1, where ``resolution'' is the related noun extracted in earlier steps. Other connected words could be subjective words like ``high'' in ``a high spatial resolution of 10 m,'' or statistical words such as ``average.'' These descriptive words provide important details about types, sentiment, and  the statistical nature of measurements. This creates opportunities for higher-fidelity grouping of like measurements (e.g. ``\textit{spatial} resolution'' versus ``resolution'') and  better profiling of trends and opportunities. For example, if a satellite's 100 m spatial resolution was described as ``insufficient'' for classifying a certain type of vegetation, this could represent an opportunity for the higher-resolution HyspIRI mission to enable new science. Extraction of these type of words is also performed using PoS labels and word dependencies, but the origin for pattern matching is the token corresponding to a related entity rather than the measurement unit. 

Marve's dependency patterns were developed heuristically by analyzing scientific literature and gradually expanded to optimize levels of precision and recall throughout the HyspIRI work and in the experiment presented here. Early experimentation suggests they generalize well to domains that don't include highly unique conventions for discussing measurements. For such edge cases, Marve's rule-based architecture is transparent, flexible, and general enough to be easily modified to identify other relationship types or patterns. 

\section{Experiments and Results}
\subsection{Data}

\begin{table}
  \caption{Experiment Data}
  \label{tab:data}
  \begin{tabular}{lccc}
    \toprule
    Source&Sentences&Measurements&Sent. w/ Measurements\\
    \midrule
    News &117&58&47\\
    Scientific &372&131&93\\
    \midrule
    Total &489&189&140\\
  \bottomrule
\end{tabular}
\end{table}

Four documents were used for experimentation: two news articles from the \textit{New York Times} and two scientific publications from the medical and remote sensing domains. The first \textit{New York Times} article, ``Dell Gets Bigger and Hewlett-Packard Gets Smaller in Separate Deals,'' was selected from the Technology section and the other, ``A Cleaning Start-Up Wielding Mops, Buckets and 700 Data Points'' was from the business section \cite{hardy_2016}\cite{zimmerman_2016}. The medical publication, ``Zika Virus Associated with Microcephaly,'' is from the \textit{New England Journal of Medicine} and the remote sensing publication, ``Satellite soil moisture for agricultural drought monitoring: Assessment of the SMOS derived Soil Water Deficit Index,'' is from \textit{Remote Sensing of Environment}\cite{mlakar2016zika}\cite{martinez2016satellite}. Although our work is centered around scientific literature, news articles provide a sense of Marve's general performance extending into non-scientific domains. The remote sensing article represents literature used in the HyspIRI work, and the medical article was selected due to the prevalence of both measurements and NLP work happening in that domain. 

For each document, individual sentences were manually extracted along with any measurement values, units, and related words contained within. The total amount of labeled sentences and measurements for each type of source is presented in Table \ref{tab:data}.  We avoided data sources with more informal language (e.g. social media) for two reasons: Marve will be most useful in domains with abundant quantitative discussion and Marve's reliance on sentence structure suggests it would perform poorly on such data.

\subsection{Setup}
The labeling of related words in the evaluation data was limited to those \textit{directly} related to the measurement, most commonly connected by a verb or nominal modifiers indicating a prepositional phrase. As an example, consider the sentence in Figure 1, ``HyspIRI has a spatial resolution of 10 m.'' In this case, ``10 m'' is directly modifying ``spatial resolution,'' which is then possessed by ``HyspIRI.'' Because there is a degree of separation between ``10 m'' and ``HyspIRI,'' Marve would only include ``spatial resolution'' as related to ``10 m'' in our experiment. Extracting second-order related words is easily achieved in Marve, but we focused on first-order relations to reduce manual labeling effort and simplify the evaluation. 

The experiment demonstrates the precision and recall of Marve's related word extraction independent of Grobid Quantities' ability to accurately extract measurement values and units. Since Marve's pattern parsing (used to identify related words) originates at measurement units, units from the ground truth data were fed to the system rather than relying on Grobid Quantities to provide these values. We assume Grobid Quantities can be incrementally improved with additional labeled values and model training. Although generating this additional training data was outside the scope of our experiment, our experience with Grobid Quantities suggests that domain-specific labels and training is necessary to achieve viable levels of precision and recall for measurement value and unit extraction. 

\begin{figure}
  \caption{An example of labeled evaluation data for a sentence containing a measurement.}
  \label{fig:eval}
\begin{lstlisting}
{
  "measurements": [
    {
      "number": "10", 
      "unit": "%", 
      "related": [ 
        {
          "Samples": [] 
        },
        {
          "formalin": ["buffered"] 
        }
      ]
    }
  ]
  "sentence_num": 41, 
  "sentence": "Samples were fixed in 10% buffered formalin 
  and embedded in paraffin."
}
\end{lstlisting}
\end{figure}
\subsection{Scoring}
Marve parses each of the 489 sentences individually. If one or more measurements are in a sentence, Marve's extraction of related words for each measurement is compared to the corresponding measurement's related words in the labeled data. Because modifiers and descriptors are relatively straightforward to extract for an already-identified related  word, they were not considered in the evaluation (e.g. ``buffered'' in Figure \ref{fig:eval}). For instances where Marve's related word extractions match the related entities in the labeled data (e.g. ``Samples'' and ``formalin'' in Figure \ref{fig:eval}), a true positive is recorded for each matched entity.  A false positive is recorded for each extracted related word without a match in the labeled evaluation data. A false negative occurs when an entity from the labeled data can't be matched to the related words extracted by Marve.

\begin{table}
  \caption{Experiment Results - Confusion Matrices}
  \label{tab:results}
  \begin{tabular}{lcc}
    \toprule
    \hspace*{3em}&Predicted Negatives&Predicted Positives\\
    \midrule
    Combined\\
    \hspace*{3em}Negatives &n/a&55\\
    \hspace*{3em}Positives &115&225\\
    Scientific&&\\
    \hspace*{3em}Negatives &n/a&36\\
    \hspace*{3em}Positives &84&143\\
    News&&\\
    \hspace*{3em}Negatives &n/a&19\\
    \hspace*{3em}Positives &31&82\\
  \bottomrule
  \linebreak
\end{tabular}
  \caption{Experiment Results - Evaluation Metrics}
  \label{tab:results}
  \begin{tabular}{lccc}
    \toprule
    Metric&News&Journal&Combined\\
    \midrule
    Precision &81.2\%&79.9\%&\textbf{80.4\%}\\
    Recall &72.6\%&63.0\%&\textbf{66.2\%}\\
    F-Score  &76.6\%&70.4\%&\textbf{72.6\%}\\
  \bottomrule
\end{tabular}
\end{table}

\subsection{Results}
Marve's precision, recall, and F-score were evaluated for the two datasets and are shown in Table \ref{tab:results}. Because Marve is rule-based system rather than a generalized statistical model, the recall metric indicates the extent to which measurement language follows concrete rules rather than how well a given model represents the data. As long as the rules generalize and are relatively concise, this type of system is attractive for its speed and transparency. Our recall results imply that a rule-based system such as Marve can identify words and entities related to measurements with high fidelity. While some recall error is expected because language is varied and often misused, these results understate Marve's recall. Similar to the findings in ClausIE's experiments, our preliminary analysis suggests that a significant portion of recall error resulted from incorrect dependency parsing rather than the occurrence of undefined patterns.

It's no surprise precision and recall were better for the \textit{New York Times} articles. Compared to scientific publications, sentence patterns in news articles are simpler. Special characters, references, diverse punctuation, and domain-specific lexicons that can fool a dependency parser are less common. Also, Stanford CoreNLP's English parser is trained on the Penn Treebank, which contains a large share of Dow Jones Newswire stories and a much smaller portion of scientific abstracts \cite{marcus1993building}.  

These results indicate that Marve is a sound approach to extracting words that compose the context around a measurement. Performance will improve as Marve's language pattern rules are further scrutinized, extended, and refined, and advances in underlying NLP approaches will also lift performance.

\section{Applications}
\subsection{Opportunities for HyspIRI}
Marve extractions enabled deeper analysis into the scientific and application-specific significance of the HyspIRI mission mentioned in section \ref{sec:motivation}. Within our corpus of remote sensing-related journal articles, we were able to extract measurement values and units with improved precision and recall. Extractions of related words and entities then allowed us to group measurement types with more confidence. They also created opportunities to link semantic publication data to other structured scientific data -- an area of research largely unexplored in Earth Science.   

For instance, extracted contextual words often contained geophysical features related to certain measurements. These included types of vegetation, soils,  minerals, rocks,  water, and man-made features, and they are often targeted for measurement by Earth Science missions where the ability to classify certain features is essential to meeting scientific objectives. One common method of classification involves analyzing a feature's reflectance, which varies across electromagnetic wavelengths to form a signature \cite{hunt1977spectral}. These signatures contain unique combinations of inflection points that enable models to distinguish between different Earth-based features.  Given their importance to mission science objectives, we were eager to explore the discussion around inflection points for certain geophysical variables. 

To achieve this, we first needed to find measurements extractions referring to the portions of the electromagnetic spectrum. This was straightforward -- measurement units of microns or nanometers were strong indicators of a spectral reference. 

\begin{figure}
\includegraphics[height=2.4in]{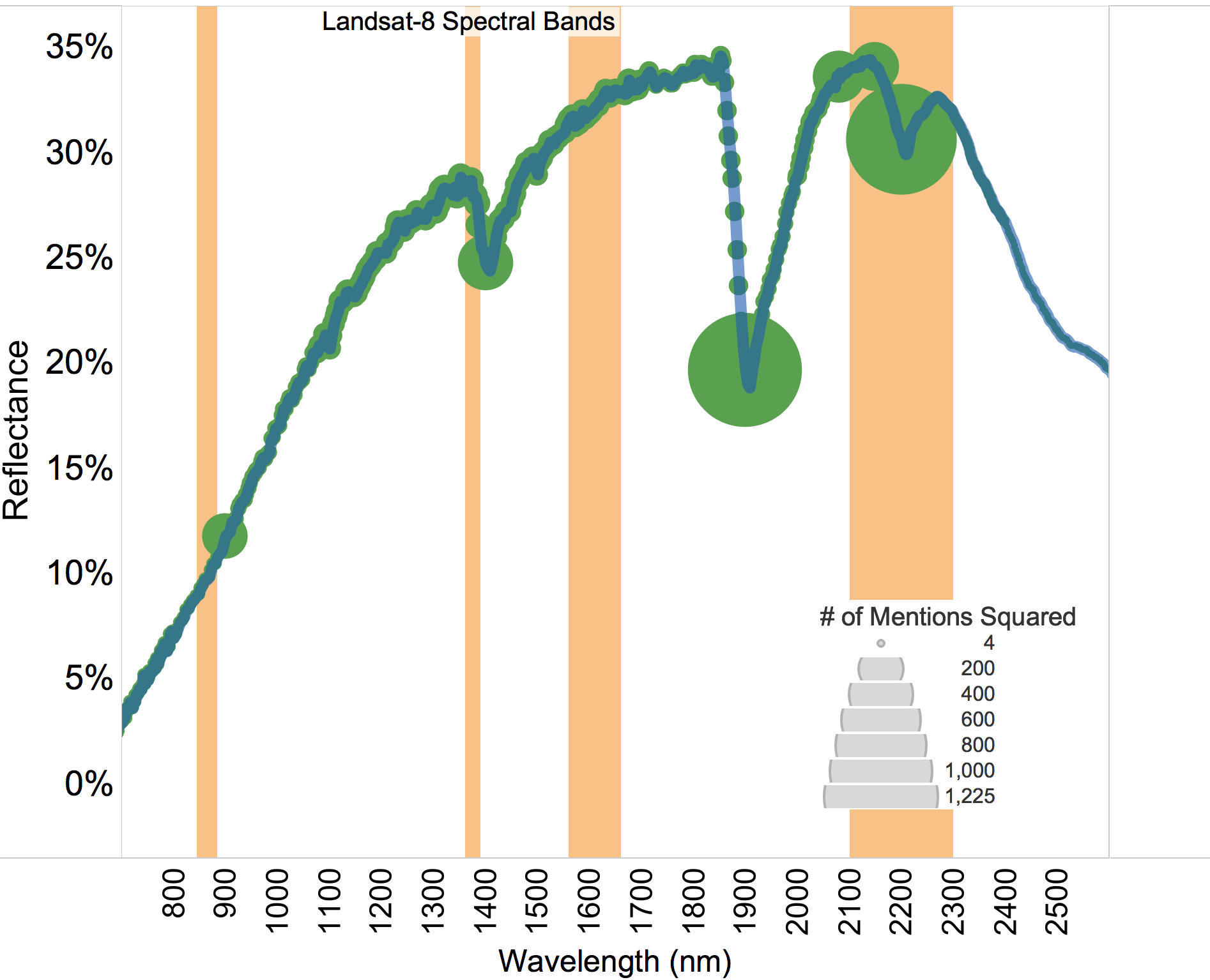}
\caption{The reflectance of a sample of brown to dark brown clay is indicated by the blue line. The green circles represent the number of extractions that indicated ``clay'' was related to a specific wavelength or range of wavelengths. The bands represent the spectral coverage of the Landsat-8 satellite. This chart supports the intuitive idea that discussion about reflectance of geophysical features would be centered around inflection points. (credit: Jason Hyon)}
\end{figure}
The next step was joining these measurements with associated reflectance signatures. The NASA Advanced Spaceborne Thermal Emission and Reflection (ASTER) mission spectral library contains over 2,300 spectra of a variety of materials, several of which were present in the related words extracted with measurements \cite{baldridge2009aster}. One such example can be seen in figure 3, which shows the reflectance signature of a sample of brown to dark brown clay. There is a large inflection point at 1,900 nm where there also are several mentions of ``clay'' in association with this wavelength (e.g. ``1900 nm'' or ``1.9 $\mu$m'' or ``1800-1900 nm''). This figure also shows mentions around wavelengths that aren't obvious inflection points. They may warrant discussion for other reasons, such as their importance in a specific scientific application.

In addition to purely scientific objectives, accurate remote sensing of clay types has tangible downstream benefits to various commercial industries. For example, specific compositions of clay mineral deposits are widely used in the production of ceramics, \cite{mbaye2012mineralogical} and clay composition information helps in understanding absorption properties, which can be used to manage water irrigation more efficiently \cite{berber2012efficient}. As previously discussed, determining the extent to which remote sensing can support these applications first involves understanding different measurement requirements. Once these are defined, assessment of existing satellites and airborne instruments is necessary to identify measurement gaps and opportunities (for HyspIRI in this case). Again, this can be accomplished through integration of structured data with Marve extractions. Figure 5 shows bands indicating the spectral coverage of Landsat-8, a satellite developed by NASA and the U.S. Geological Survey (USGS) and launched in 2013 \cite{Landsat8} \cite{NASA}. The pronounced inflection point at 1,900 nm isn't captured by Landsat-8, which could represent an opportunity for HyspIRI. If extended analysis reveals that publicly-sponsored satellites are missing key measurements, these gaps can inform decisions about HyspIRI's future resolution and spectral coverage capabilities. In this sense, HyspIRI would be uniquely positioned to support key priorities as defined by the broader scientific community.      

\subsection{Knowledge Base Construction}

The construction of a measurement knowledge base for remote sensing publications would be of great value to researchers and NASA administrators. For example, in the \textit{Remote Sensing of Environment} paper used in our experiment the authors write, ``Although there is more and more information about these soil water parameters, they are not usually included in standard soil databases. For that reason, researchers sometimes have simply used soil parameter data published in the literature.'' A structured repository of measurement information would allow researchers to better explore scientific results, experimental designs, instrument specifications, and general discussion around specific measurement types and their relationships to time, locations, organizations, and other domain-specific entities. The value and feasibility of constructing similar knowledge bases has been demonstrated by several projects including our own prior work in constructing a Polar Cyberinfrastructure that integrates measurements crawled from the U.S. National Science Foundation Advanced Cooperative Arctic Data and Information System (ACADIS), the NASA Antarctic Master Directory (AMD), and the National Snow and Ice Data Center (NSIDC) Search tool as part of constructing the National Institutes of Standards and Technology Text Retrieval Conference (TREC)-Dynamic Domain (TREC-DD) polar dataset\cite{burgess2015trec}. In addition, Stanford's DeepDive platform was used to construct PaleoDeepDive, a system that has processed 300,000 scientific documents in an effort to replicate the manually curated Paleobiology Database (PBDB). This database contains hundreds of thousands of taxonomic fossil names and attributes manually entered by researchers over the last two decades. PaleoDeepDive recreated PBDB with greater than double human and roughly equal precision. 

Marve significantly reduces the manual effort needed to create such a knowledge base and can be easily integrated into DeepDive, which can add non-measurement-based extractions and also help categorize measurements and their relationships with other entities. DeepDive has previously been used to derive relationships between measurements and related words or entities (e.g. PaleoDeepDive), but users are left to their own devices to extract measurements and possible related words and entities. They also need to write features used by the inference engine to identify and classify relationships between extractions. Marve automates these extractions and can provide DeepDive developers with valuable features in the form of measurement context.
\section{Future Work}
\subsection{Expanding Experimentation}
Research and applications involving measurement extraction are limited. Marve creates second-order extractions (SOE) constructed from other first-order extractions (FOE) that are either new (Grobid Quantities) or have progressed significantly in recent years (word dependencies, PoS tagging). As a result, publicly-available labeled datasets do not exist for evaluating Marve. Our evaluation dataset is relatively small, and we hope to extend this dataset and employ a linguist for review.

Marve is also tightly coupled with CoreNLP and CoreNLP errors could not be isolated in the experiment without hand labeling of POS tags and word dependencies. This process is tedious and requires lingual expertise to perform accurately and consistently. While we have focused on measurement-rich scientific publications in our development and applications of Marve, we plan to explore its performance on syntactically labeled data such as the Penn Treebank \cite{taylor2003penn}. Although measurements are sparser, experiments with such data sources allow for Marve to be evaluated independent of FOE.   

Lastly, we are also interested in understanding the performance of Grobid Quantities at different levels of training and customization. Generating additional training data was outside the scope of our experiment, but we are curious how well a Grobid Quantities model trained on a pre-existing set of diverse labeled data would generalize. This will allow practitioners to weigh the costs and benefits of domain-specific labels and training by understanding Grobid Quantities' off-the-shelf performance on unseen data.

\subsection{Extending Marve}
Expanding the semantic information embedded in Marve extractions increases the potential for automatic classification of measurements. While Marve represents a large step forward in the collection of this information, the burden is on the user to make use of it, which will involve grouping or classifying measurements in almost all cases. DeepDive addresses this problem by allowing users to write ``labeling functions,'' which provide the system with features used to classify different types of relationships. We plan to explore further integration of Marve into DeepDive and its new successor, Snorkel, while also exploring unsupervised approaches to measurement grouping. We view providing a means for automatically or semi-automatically classifying measurements as an important step in Marve's development.  

\section{Conclusion}
We propose a baseline method, Marve, for contextual measurement extraction, a sub-area of information extraction that has been largely unaddressed in the research community. Semantic measurement information is inherently richer than raw data, and our initial findings with Marve are positive. As the world becomes increasingly inundated with textual data, Marve and other related approaches will help us find relevant scientific information and develop a broader understanding of our domains. We view Marve as an opportunity to expedite scientific research and inform scientific investment, two areas essential to encouraging innovation and demonstrating the importance of science to society.

\begin{acks}
This effort was supported in part by JPL, managed by the California Institute of Technology on behalf of NASA, and additionally in part by the DARPA Memex/XDATA/D3M programs and NSF award numbers ICER-1639753, PLR-1348450 and PLR-144562 funded a portion of the work. The authors thank Jason Hyon, Paul Ramirez, Drs. David Thompson, Diane Evans, Randy Friedl, and Duane Waliser for their support and feedback.
\end{acks}

\bibliographystyle{ACM-Reference-Format}
\bibliography{KDD17-marve} 

\end{document}